\documentclass[11pt]{amsart}
\usepackage{geometry} 
\usepackage{color}
\usepackage{geometry}
\usepackage{amssymb}
\usepackage{graphicx}
\usepackage{graphics}
\title{\small Three-dimensional dynamics 
of a fermionic Mott wedding-cake in clean and disordered optical lattices}
\date{} 

\begin{document}

\maketitle
\begin{center}
\textsc{A. Kartsev, D. Karlsson, A. Privitera and C. Verdozzi}
\end{center}

\begin{abstract}
{\bf Non-equilibrium quantum phenomena are ubiquitous in nature. Yet, theoretical predictions on the real-time dynamics of many-body quantum systems remain formidably challenging, especially for high dimensions, strong interactions or disordered samples. Here we consider a notable paradigm of strongly correlated Fermi systems, the Mott phase of the Hubbard model, in a setup resembling ultracold-gases experiments. We study the three-dimensional expansion of a cloud into an optical lattice after removing the confining potential. We use time-dependent density-functional theory combined with dynamical mean-field theory, considering interactions below and above the Mott threshold, as well as disorder effects. At strong coupling, we observe multiple timescales in the melting of the Mott wedding-cake structure, as the Mott plateau persist orders of magnitude longer than the band insulating core. We also show that disorder destabilises the Mott plateau and that, compared to a clean setup, localisation can decrease, 
creating an interesting dynamic crossover  during the expansion. }
\end{abstract}

\section*{Introduction}

A large part of our understanding of the physical world concerns the equilibrium state, where macroscopical observables are constant in time. However, most 
phenomena surrounding us are instead non-equilibrium phenomena, where systems evolve towards new equilibrium states once perturbed away from their initial conditions. For condensed matter systems, 
ultrafast techniques are now becoming available to probe the non-equilibrium dynamics at the typical electronic scales \cite{fausti_science,giannetti_science}. 
Even so, the intrinsic complexity of real materials still makes their theoretical understanding extremely difficult, because the electronic effects are usually 
interwoven with phononic contributions, inhomogeneities,  etc.

In contrast, idealised solids realised by loading fermionic ultracold atoms into optical lattices feature several advantages over
the corresponding solid-state systems. The accurate control of the system parameters permits experimental realisations of model 
fermionic Hamiltonians, e.g. the single-band Hubbard model \cite{Hesslinger,Rigol2D}, where higher-bands contributions, phonons, lattice defects are absent 
or can be introduced separately and in a controlled way. In fact, most system parameters can be tuned independently in time and 
in a very wide range, allowing for the exploration of regimes beyond what is feasible in condensed matter \cite{BlochRevModPhys08,bloch_nature_review}. Moreover, 
the typical timescales of ultracold-gases dynamics
are much longer ($ \approx ms$) than those in condensed matter systems \cite{Rigol11}. This has made possible, for example,  the observation of interesting phenomena like the crossover from ballistic to diffusive behaviour in a fermionic cloud expanding into an optical lattice \cite{bloch_nature}.
Although the physics in these experiments is much simpler than in condensed-matter ones, the
theoretical understanding is still in general impaired by the absence of reliable tools to address the non-equilibrium behaviour of these simple model fermionic systems.

Indeed, among the well established approaches, perturbative techniques
are only suitable for weak- and strong-coupling regimes, while non-perturbative schemes scale unfavourably with system size and simulation time. On the whole, it is fair to say that using any of the methods above for large inhomogeneous samples does not appear immediate, because of rather prohibitive Êcomputational costs.
Lastly, exact techniques like the time-dependent density-matrix renormalisation group are currently mainly restricted to one dimension, while Quantum Monte-Carlo
approaches can be severely limited by the fermionic sign-problem. Therefore, progress in the field requires new methods to cope with higher dimensions, large inhomogeneous systems irrespective of the interaction strength. 

For these challenging situations, we advocate here the use of 
a recently developed approach \cite{KarlssonPRL11}, i.e. the combination of Lattice Time-Dependent Density-Functional Theory and Dynamical Mean-Field Theory  (Methods).  Within this method, one can properly describe the Mott transition within a density-functional theory framework  \cite{KarlssonPRL11}. 
Here, we show its potential by  addressing the real-time dynamics of a large inhomogeneous three-dimensional system (up to $47^3$ lattice sites)
in different regimes of interaction strengths and also in presence of disorder. 

Our setup resembles a recent experiment\cite{bloch_nature} on ultracold Fermi gases, consisting of a confined cloud, 
which expands into an optical lattice after the trap is removed.  Our initial density profile is the ground-state density of the trapped interacting cloud. 
The corresponding initial state is the (Kohn-Sham) ground state  of density-functional theory (Methods).
To describe the expansion of the system, we study the time-dependent one-particle density, which can
 in principle be obtained exactly within our approach (Methods).

We considered two interaction regimes, corresponding to a strongly-correlated metal and to a Mott insulator in the homogeneous case, 
and different protocols for switching off the trapping potential. At strong coupling, the ground-state density exhibits the peculiar wedding-cake
 structure due to the presence of the {\emph{Mott plateau}} where the local density is commensurate \cite{weddingcake}. 
As we release the trapping potential,  the high-density metallic domain immediately melts, while the Mott plateau remains remarkably
stable against the expansion, over much longer timescales than below the Mott transition. Thus the intrinsic inhomogeneity 
of confined ultracold gases displays the multi-scale dynamics of different phases in a single experiment. 

We also considered the role of disorder on the cloud expansion.  Compared to the static case\cite{Abrahams10,Inguscio07,Polinidisord,Gaodisord,NatureLocalization12}, the effect of disorder on interacting Fermi systems out-of-equilibrium is much less understood, especially 
for dimensions larger than one. Our results show
that disorder earlier makes the Mott plateau less stable, decreasing the melting time, while slowing down the expansion at long times. This induces a noteworthy crossover in real-time. 

As commonly done within the ultracold-atom community, the system is described in terms of an inhomogeneous and time-dependent Hubbard model. In standard notation,
\begin{equation}
 {\mathcal H}=-J \sum_{<i,j>,\sigma} c^\dagger_{i,\sigma}c_{j,\sigma} + U \sum_{\mathrm i} n_{\mathrm i\uparrow} n_{\mathrm i\downarrow} + \sum_{i,\sigma}  (V_0(t) r^2_i -\varepsilon_i) n_{i,\sigma}
\label{hamiltonian}
\end{equation}
with $J$ the hopping parameter and $U$ the contact interaction strength.
The trap strength $V_0(t)$ determines  the switching-off protocol in time, and $\varepsilon_i$ is nonzero in
the disordered case (Methods). We take $J$ as energy unit, thus $t$ is in units of $\hbar / J$.

In our simulations, we consider a simple-cubic lattice with $U=8$ and $U=24$. In the homogenous case, these values 
correspond respectively to a strongly-correlated metal and to a Mott insulator (the critical interaction $U_c \approx 13 $).
Starting from the trapped system in the ground state,
we examine three different expansion scenarios: For $U=8$, we study a \emph{sudden} expansion ($\tau=0^+$), as well as a \emph{slow} one ($\tau=80$), where $\tau$ controls the trap removal speed (Methods). For $U=24$, instead, we choose to consider only the slow expansion. 
 
Indeed, the current implementation of our approach  is expected to give a reliable description for slow and moderately-fast expansions, even at strong coupling.
However, for fast trap removals, its performance significantly deteriorates for increasing interaction, and large deviations occur well above the Mott (gapped) regime.
Instead, in the metallic (gapless) regime at $U=8$, our results should remain robust in a qualitative sense. This is further addressed in the Methods section.

\begin{figure}[tbh]
\includegraphics[width=13cm]{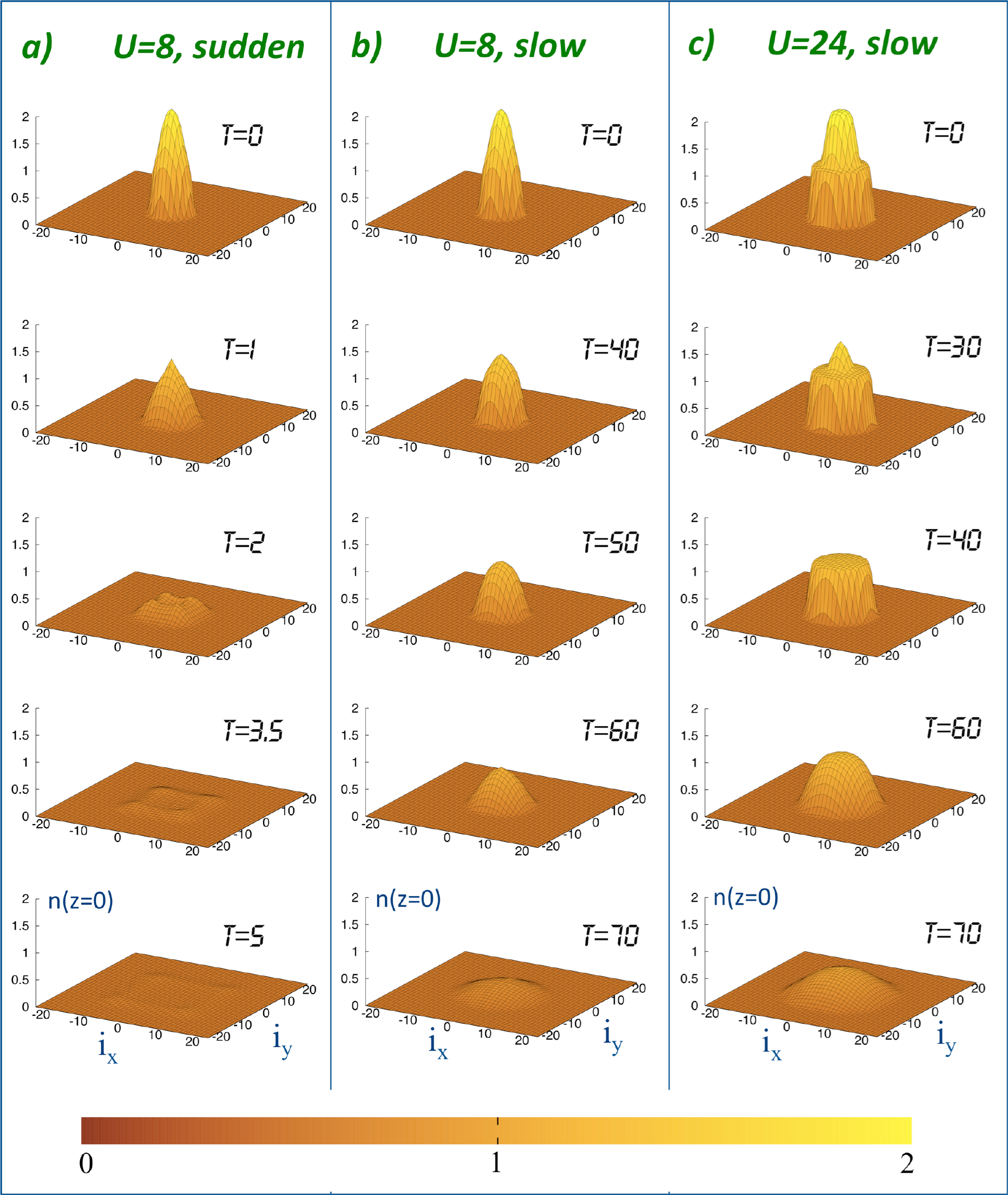}
\caption{ {\bf  Cloud expansion into a homogeneous lattice.}
{\bf a},
{\bf b},
{\bf c}: Density profiles  along the $z=0$ plane for different times and setups. The colour bar represents the density scale.}
\label{Fig.1}
\end{figure}

\section*{Results}
{\bf Expansion into a homogeneous lattice. $\!$} The qualitative behaviour in time of the density profile in the $z=0$ plane for the chosen setups is shown in Fig. 1.
Even within a cubic lattice, the initial cloud profiles appear to be in very good approximation spherical, due to the symmetry of the trapping potential. In Fig. 1a-b, for $U=8$, the initial density profile smoothly decreases as a function of the distance from the centre in every direction. At strong-coupling instead ($U=24$, Fig. 1c), the repulsion is large enough to induce an insulating phase in the homogeneous system at half-filling.  Accordingly, the trapped system develops a Mott plateau, i.e. a region where the density is $n=1$
to a very good approximation, due to the incompressible nature of the Mott phase.

\begin{figure}[tbh]
\includegraphics[width=14cm]{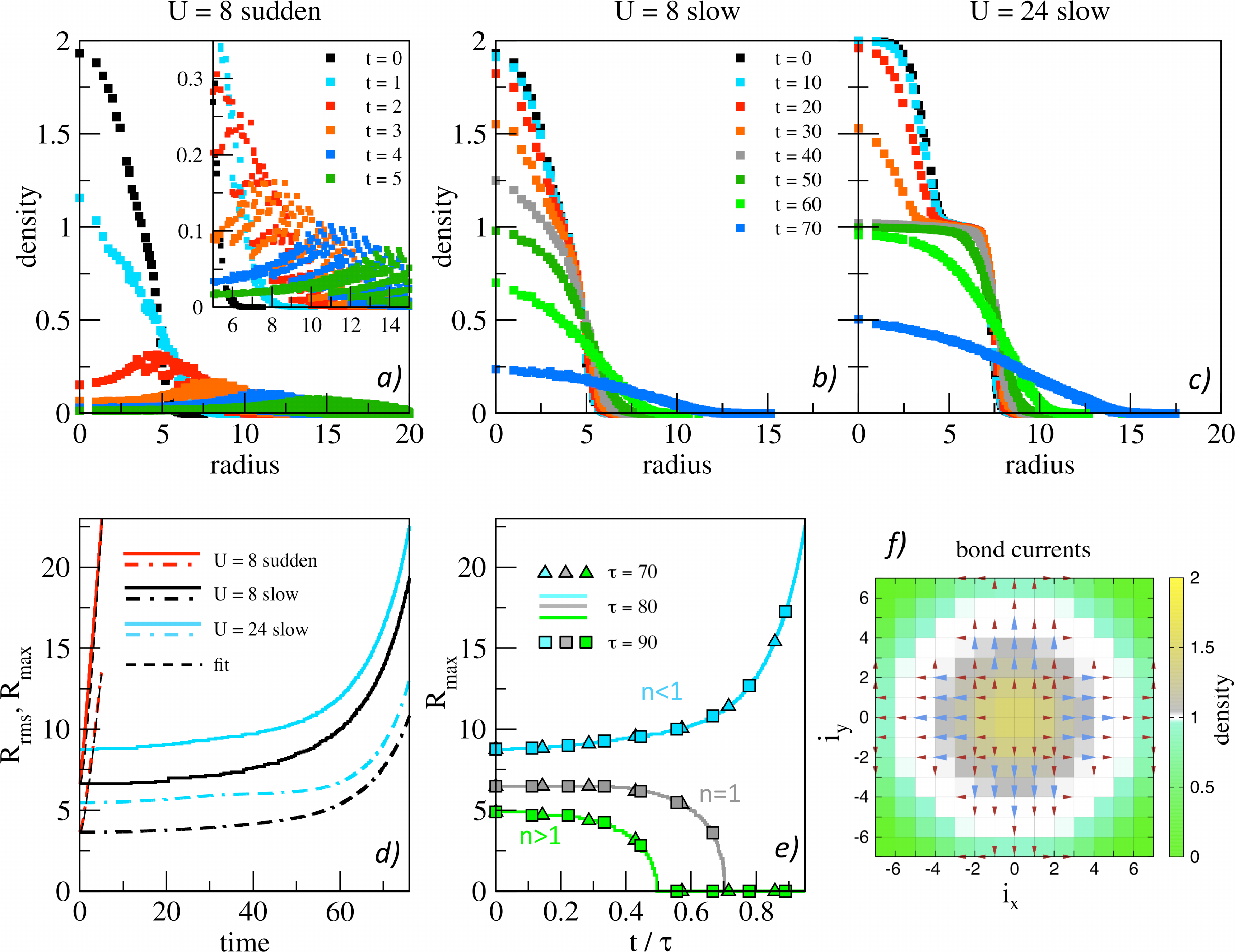}
\caption{{\bf Quantitative analysis of the homogeneous expansion for different times and setups.}
{\bf a},
{\bf b},
{\bf c}: Density as a function of radius only. The magnification in ({\bf a}) details the cloud anisotropy. The inset legend also applies to the rest of ({\bf a}). The legend in ({\bf b}) is shared with ({\bf c}).
{\bf d}: $R_{max}$ (solid lines) and $R_{rms}$ (dashed-dotted lines). For $U=8$, sudden expansion, fits are shown (dashed lines).
{\bf e}: $R_{max}$ for $U=24$ as a function of rescaled time $t/\tau$. Densities for different trap protocols ($\tau=70,80,90$) have distinct symbols; different density domains have distinct colours.  {\bf f}: Bond currents and densities in the $z=0$ plane for $U=24$, $t=30$. Blue (red) arrows correspond to large (intermediate) currents, whilst small currents are not shown.}
\label{Fig.2}
\end{figure}

It is immediately manifest that the Mott physics has striking effects on the dynamics. Indeed, for $U=8$ the cloud expands smoothly (Fig. 1a-b), while for $U=24$ (Fig. 1c) the band insulating core immediately melts to get rid of the large interaction energy, and the upper part of the wedding-cake structure at $n > 1$ collapses over the Mott plateau. At the same time, the underlying Mott plateau is remarkably stable against the trap opening and, after some time, the system exhibits a large Mott region surrounded by a metallic domain (Fig. 1c, $t=40$). Only on a much larger timescale ($t > 60$), also the Mott plateau melts and the system fully relaxes into the lattice.

Different switching-off protocols are compared in Fig 1a-b. For the sudden case, the system evolves only according to the homogeneous Hamiltonian, which (together with the initial state) sets the expansion timescales.  As a consequence, the metallic dome melts much faster, and the shape of the cloud develops clear signatures of the lattice symmetry ($t>2$). Conversely, in the \emph{slow} case, the expansion rate is controlled by the trap-removal speed, 
and thus the cloud anisotropy remains small.

In Fig. 2, we analyse the expanding cloud in a more quantitative way.
We start by looking at the density as a function only of the radius $r=\sqrt{i_x^2+i_j^2+i_z^2}$ (Fig. 2a-c).
For the {\it slow} expansion (Fig. 2b-c), at $t=0$ the cloud core is basically spherical (i.e. the density is a single-valued function of $r$) and tends to maintain this symmetry. Significant anisotropy is instead observed in the low-density region around the cloud boundaries for the \emph{sudden} setup (Fig. 2a).  In this case one expects the tail expansion to be ballistic \cite{bloch_nature}, with the cubic symmetry of the underlying lattice.  

These considerations are further supported by the analysis (not shown) of an Óasymmetry parameterÓ $\gamma=(R_{max}-R_{min})/(R_{max}+R_{min})$, where $R_{max}$ and $R_{min}$ are the maximum and minimum radial size of the cloud ($\gamma \in [0,1]$). The cloud radius $R_{max} (R_{min})$ is defined as 
the largest (shortest) distance from the centre where the density is above $10^{-4}$ ($R_{max}$ was also used to check that the expanding cloud did not reach the boundaries
of the simulation box).  The maximum asymmetry we found is $\gamma=0.38$ for the sudden case, and $\gamma = 0.05$ for the slow expansions.
Finally, a notable feature in Fig. 2c is that the Mott plateau, before collapsing, widens over a significant time interval ( $0 \lesssim t \lesssim 40$). 
This appears to be independent of dimensionality  \cite{DKCVMKC11}, and thus related to the intrinsic energetics of the Mott phase and its rigidity.

Another useful quantity to analyse is the cloud expansion velocity. This was done in ref. \cite{bloch_nature}, where, starting from an initial band-insulator state, the dual nature of the expansion was characterised as ballistic at the cloud edge, and, due to interactions, as hydrodynamic in the cloud core.  In our analysis, we compare the maximum $R_{max}$ and average radius of the cloud $R_{rms}=\sqrt{\frac{1}{N_p} \sum_i n_i r_i^2}$, where $N_p$ is the number of particles. Their different behaviour (Fig. 2d) can give indications on the presence of multiple domains expanding at different rates.  For $U = 8$ and in the case of a sudden expansion, $R_{max}$ and $R_{rms}$ were fitted according to the expression $ \sqrt{R_0^2 + v^2 t^2}$, where $R_0$ generically denotes   $R_{max}$ ($R_{rms}$) at $t=0$, and the speed $v$ is a fitting parameter. The fit of $R_{max}$ confirms that the edge of the cloud expands ballistically \cite{bloch_nature}.
At the same time, taking into account the slower interacting core, $R_{rms}$ increases at a lower rate than $R_{max}$.
For the slow expansions, also shown in Fig. 2d,  no fit was attempted, since the the slow trap opening hinders the expected ballistic behaviour at the cloud tails.

The effect of different trap-removal protocols on the dynamics in the Mott regime is shown in Fig. 2e for $U=24$.
We find it informative to scale time according to $t \to t / \tau$. 
In this way, simulations for different $\tau$:s appear very similar, and can be discussed together. This also suggests that the intrinsic cloud relaxation timescales 
are much faster than $\tau$.  Furthermore, due to the presence of the Mott plateau, we find it insightful to describe the cloud as consisting of three domains, naturally defined in terms of the density: a low-density  ($n < 1$) and high-density  ($n>1$) metallic region and the Mott plateau ($n = 1$).  For each of them, we consider a distinct  $R_{max}$
and the related expansion speed.  For $n<1$, we find a positive speed (for $\tau=80$, the low-density region $R_{max}$ is the same as the original $R_{max}$ in Fig. 2d). Nevertheless, the speed is negative for the other domains, which means that their outer radius is shrinking in time (note, however, that the Mott plateau is growing inwards). Even after the high-density core has disappeared, the Mott region has only just started to shrink, thus confirming that the different domains have qualitative different behaviours.

Finally, in Fig. 2f  we present results for the Kohn-Sham  bond-currents at time $t=30$. Although their physical meaning is rather indirect,  
such currents can offer additional insight (Methods). From their intensity and direction, we see that the particles are flowing out of the 
high-density region in the centre. However, no particles accumulate in the Mott region, consistently with the rigidity of the latter.

The foregoing discussion for a clean lattice reveals several interesting dynamical features of strongly correlated
fermions. Another important element that we wish to bring into our analysis is the effect of disorder, which we discuss next.

\begin{figure}[tbh]
\includegraphics[width=14cm]{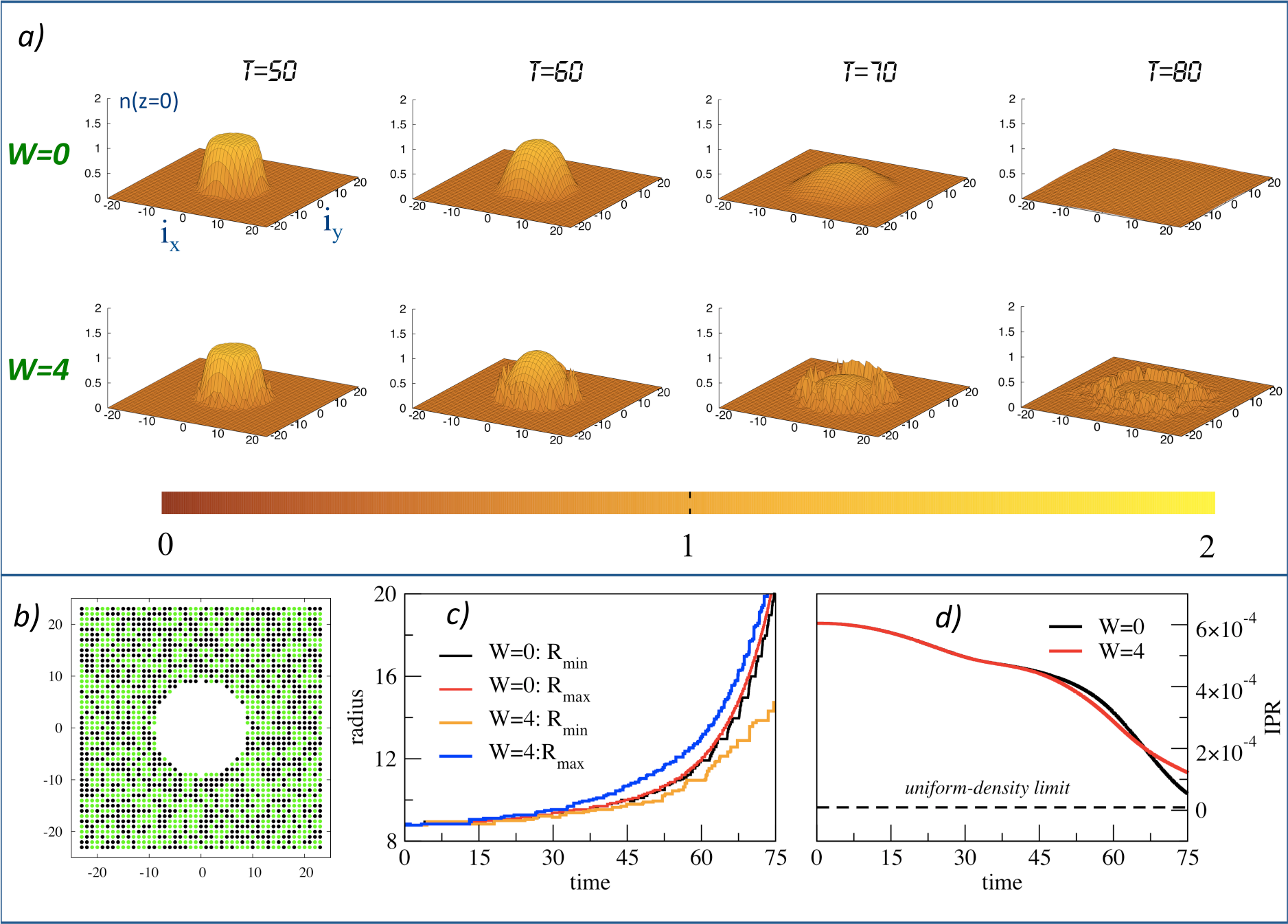}
\caption{{\bf  Mott wedding-cake expansion in clean and disordered systems.}
{\bf a}: Comparison between the density profiles in the $z = 0$ plane for the disordered $(W = 4)$ and clean $W=0$ (same as in Fig 1c) systems. 
{\bf b}: $z=0$ section of the used special quasi-random structure.
{\bf c}: Relative trends between the clean and disordered expansions for the minimum and the maximum cloud radii.
{\bf d}: Time evolution of the inverse participation ratio (IPR) for the clean and disordered expansions.
}
\label{Fig.3}
\end{figure}

\noindent {\bf Expansion into an inhomogeneous lattice. $\!$} To illustrate the role of disorder, we consider one particularly interesting situation,
namely the dynamical competition between disorder and interaction in the Mott regime.
To this end, we prepared the system for $U=24$ in the same initial state as in the previous section and let it expand into a disordered environment
according to the \emph{slow} trap-opening protocol used before. Disorder was introduced via a single \emph{special quasi-random} structure (Methods), which mimics a 50\% binary alloy ($\varepsilon_i =\pm W$ in Eq. \ref{hamiltonian}). The $\varepsilon_i$:s, as shown in Fig. 3b, are chosen to be non-zero only where the initial density is negligible, 
and thus we can specifically address the role of disorder on the cloud dynamics.

In Fig 3a, we show the density profile along the $z=0$ plane for the homogeneous ($W=0$) and the disordered case $(W=4)$. 
For this setup, the effect of disorder on the density profile is hardly visible on the scale of the figure for  $t \lesssim 50$, due to the peculiar Mott plateau dynamics.  
This is easily understood by noting that, initially, the expansion is mostly characterised by an internal rearrangement of the high-density region $n \geq 1$
and the formation of the extended Mott plateau, as in the clean case. This has a relatively small effect on the low-density region. 

The influence of disorder is instead much more pronounced for $t>50$ when the Mott plateau begins to collapse and the particles around its boundaries flow towards the low-density region. A first more evident effect is that the  cloud expansion at large times is now hindered by the scattering against the binary disorder, which causes an irregular density accumulation close to the inner edge of the disordered region. At large times $t \geq 70$, the disorder induces a kind of dynamical localisation of the cloud as the density profile is significantly reduced in the trap centre 
(in fact more than in the $W=0$ case) but at the same time the expansion far away from the centre is considerably slowed down.
The final density profile at $t=80$ shows a large particle accumulation in a roughly annular region at the beginning of the disordered zone, in contrast to the clean case, which is almost uniform (Fig. 3a). A second and more subtle effect, hardly visible on the plot scale and discussed further below, is that the disorder accelerates the melting of the Mott plateau.

To quantify the effect of disorder on the size and symmetry of the expanding cloud, we also analysed    
both the minimum $R_{min}$ and maximum $R_{max}$ radial size of the cloud with and without disorder. Not unexpectedly, in presence of disorder $R_{min}$ and $R_{max}$ rapidly diverge from each other, since disorder destroys the spherical symmetry observed in the clean case.
$R_{max}$  separates from its  clean-setup counterpart in the first stages of the cloud expansion,
indicating that disorder earlier favours some particles to flow away from the cloud in a faster way. Interestingly,
disorder induces a larger and more asymmetric cloud at every step of the expansion.

The inverse participation ratio $\zeta$,  shown in Fig. 3d for $W=0$ and $W=4$ as a function of time,  is a convenient indicator of the degree 
of localisation resulting from competing disorder and interactions (Methods). Since a smaller $\zeta$ implies a smaller degree of localisation for the particle cloud, in the first stage
 of the expansion ($t \lesssim 65$) we observe a rather interesting and counterintuitive result:  for $W=4$, $\zeta$ stays smaller than for the homogeneous case (whereas disorder in general is expected to increase the degree of localisation), resulting in a more delocalised cloud at small expansion times.
The dynamical localisation in Fig. 3a takes over  at large times and the trend is inverted for $t>65$. 

This dynamical crossover is an intriguing consequence of the competition between disorder and interaction in the Mott regime. 
Indeed the disordered lattice offers in the early stages of the expansion additional energetic pathways compared to the $W=0$ case, which accelerate the melting  of the Mott plateau. This is also why the expansion initially occurs earlier and faster than in the clean case (thus we observe a smaller  $\zeta$ and a larger $R_{max}$). 
However, after the Mott plateau has melted, the cloud expands faster (on average) in the clean setup, since the particle flow is not hindered by disorder 
(hence the larger $\zeta$ at $t \gtrsim 70$ for the disordered case). We verified that the overall features are robust against distinct quasi-random structures.  

\section*{Discussion}

We have studied the dynamics of an interacting fermionic cloud expanding into clean and disordered optical lattices by using a recently developed approach which merges lattice Time-Dependent Density-Functional Theory with Dynamical Mean-Field Theory. With this method, we have been able to describe the real-time dynamics of large three-dimensional (in)homogeneous Fermi systems up to an unprecedented size ($N \sim 10^5$ lattice sites, i.e. comparable with current experiments in cold gases), in different regimes of interaction and even in presence of disorder. Our work unveils important aspects of the fermionic dynamics, such as the central role of the Mott physics at strong-coupling and the interplay between disorder and interaction in 3D. 

Above the Mott threshold, the timescales and features of the clean expansion are markedly different from the metallic regime. We observe an earlier increase in size of the Mott plateau against the trap opening at the expense of the metallic domain. Compared to lower dimensions, this finding is even more surprising, due to the larger number of runaway paths in 3D:
this mainly arises from the universal features of Mott physics. The 3D nature of the system is instead crucial i) in the observed dichotomy between the weak- and strong-coupling dynamics, which reflects the presence of a finite Mott threshold in the homogeneous system. Our findings also suggest a convenient description of the cloud expansion in terms of multiple domains with positive and negative expansion speeds ii) in the rich phenomenology observed in presence of disorder.  

Disorder introduces notable changes in the dynamics above the Mott threshold, altering expansion time-scales, but also resulting in interesting temporal patterns. For example, we observe a dynamic crossover in the localisation properties,  as the disordered system is less localised than the clean one at the beginning of the expansion, whilst disorder increases in general localisation, as expected, at large times.

Our results shown here are just an example of how rich and interesting the real-time dynamics in the  
presence of disorder and interactions is. We have already considered (not shown) different disorder strengths and interactions,  and  observed intriguing features like, e.g., density revivals due to dynamical localisation.  These findings, as well as  the inclusion of proper configuration averaging and disorder in the initial state, are deferred to future publications.

In conclusion, we hope our work will stimulate further improvements of our approach, as well as new experiments on ultracold gases in optical lattices, to deepen the understanding of many-body quantum systems in many regimes of interest. 

\section*{methods}
To describe the properties of the inhomogeneous Hubbard model in Eq. (\ref{hamiltonian}),
 we use static \cite{HK64,KS65} and Time-Dependent  \cite{RG84}
Density-Functional Theory (DFT and TDDFT, respectively). These   
are in-principle-exact reformulations of the many-body problem, and we use them 
in their (static \cite{GS86,GSN95,Lima03,KlausRigol,sanvito10} and time-dependent
\cite{Ferdi,Magyar,Verdozzi08, Baer08,Polini08,GSEPMC10,KurthSte11,DKCVMKC11,KarlssonPRL11,CVChem11,Tokatly12}) lattice versions.

Here, we briefly recall the essentials of our treatment (for a recent review, see Ref. \cite{CVChem11}).
In this approach, the (time-dependent) number of particles per lattice site  $n$,
is the basic variable and the physical observables of the system are functionals of $n$. 
In operational terms, one introduces a non-interacting image system, the so-called Kohn-Sham (KS) system,
and the exact many-body density is then obtained
from the KS single-particle states.  A key ingredient in the KS system
is the exchange-correlation potential, $v_{xc}$, incorporating exchange and correlation effects. In general, $v_{xc}$ is not known exactly and approximations are used.
A simple but effective one (used here, and further discussed below) is the Local Density Approximation (LDA) for the static case, 
where $v_{xc}$ at site $i$ depends locally on the site occupation $n(i)$, and correspondingly 
the adiabatic LDA (ALDA) for the time-dependent case, where $v_{xc}$
depends instantaneously and locally on $n(i,t)$. 

In a recent work, a lattice DFT treatment of simple-cubic Hubbard model\cite{KarlssonPRL11} was proposed,
where the pivotal ingredient is an adiabatic LDA based on Dynamical Mean-Field Theory (DMFT) \cite{MetzVoll89,RevModPhys96,KoVo2004}. 
There, $v^{hom}_{xc}$ for the reference homogeneous system was obtained within DMFT according to
 \begin{eqnarray}
 v^{hom}_{xc}(n)= \frac{\partial}{\partial n} ( E_{DMFT} (n) -T_0 (n)- Un^2/4),
 \end{eqnarray} 
where $E_{DMFT}$ is the ground-state energy, $T_0$ is the non-interacting kinetic energy and $Un^2/4$ is the Hartree energy.
DMFT properly describes the Mott metal-insulator transition \cite{RevModPhys96},
and gives a good variational estimate of the energy \cite{prokofev}, although
the self-energy only depends on the frequency and not on quasi-momentum.

A crucial feature of the method is the occurrence of a discontinuity in $v_{xc}$  at  $n=1$ above a critical value $U_{c}  \approx 13$
of the Hubbard interaction. This is how the Mott-Hubbard metal-insulator transition manifests within 
a DFT framework. The discontinuity is the origin of the Mott plateaus in Fig.\ref{Fig.1}, for $U=24 > U_c$. 

In our calculations, we considered simple-cubic clusters of  $47\times 47 \times 47$ lattice sites with open boundary conditions. We chose $N_p=542$ $(N_p=1862)$ particles when $U=8$ $(U=24)$ to avoid ground state degeneracies in the density region of interest. The ground state was computed by solving 
self-consistently the KS equations:
\begin{eqnarray}
(\hat{T}+\hat{v}_{KS})\varphi_k=\epsilon_k \varphi_k, \label{KSDFT}
\end{eqnarray}
where the effective potential ($i$ labels the lattice site) $v_{KS}(i)=v_{ext}(i)+Un(i)/2+v_{xc}(i)$, with
$\hat{T}$ the kinetic energy operator on the lattice, and
$Un(i)/2$ the Hartree contribution, with $n(i)=2\sum_{k \in occ} |\varphi_k(i)|^2$ (the sum is over all occupied KS orbitals, and
the factor 2 accounts for spin degeneracy). $v_{ext}(i)=V_0 (t < 0) = Kr_i^2$ is the external trapping potential. 
We chose $K = 0.55$ $(K=0.60)$ for $U=8$ $(U=24)$. In the LDA, $v_{xc}(i) = v^{hom}_{xc}(n(i))$.
The large-scale self-consistent eigenvalue problem of Eq. (\ref{KSDFT}) was made computationally manageable
by using symmetry-adapted orbitals via the point group $O_h$.  
Also, for $U=24$, $v_{xc}$ was slightly smoothened
for numerical convenience (see e.g. ref. \cite{DKCVMKC11}). 
As a result, the Mott domain is not perfectly flat. 
The threshold for self-consistency in the density was $\leq10^{-5}$.

On removal of the trap, the system was evolved in time via the time-dependent
KS equations 
\begin{eqnarray}
(\hat{T}+\hat{v}_{KS}(t))\varphi_k(t) =i\partial_t \varphi_k(t), \label{KSTDDFT}
\end{eqnarray}
where $v_{KS}(i,t)=v_{ext}(i,t)+Un(i,t)/2+v_{xc}(i,t)$, with (in the ALDA)
$v_{xc}(i,t)=v^{hom}_{xc}(n(i,t))$,
and $n(i,t)=2\sum_{k \in occ} |\varphi_k(i,t)|^2$. 
The time-dependent external potential is
 $v_{ext}(i,t)=v_{trap}(i,t)-\varepsilon_i$. When present, the disorder $\varepsilon_i$ is static,
whilst the trap is removed according to $v_{trap}(i,t)=V_0(t)r_i^2$.
We chose $V_0(t)=K\cos^2(\frac{\pi t}{2\tau})$ for $t\le \tau$ and $V_0(t)=0$ when $t >\tau$ ensuring a smooth time dependence.
The numerical time propagation was performed via the short iterated Lanczos algorithm \cite{Park86},
with a time-step $\Delta t=0.01$.

The (A)LDA neglects memory effects and the (dynamical) broadening of the discontinuity in $v_{xc}$ for inhomogeneous systems (in general, these limitations become more severe at lower dimensions). As shown by recent benchmarks ,\cite{Verdozzi08, Polini08, PUIG10PRB, CVChem11, DKCVMKC11,AnnaMaya11}, where non-locality in time and space is fully taken into account, the (A)LDA performs poorly for strong interactions (e.g. in gapped Mott systems) when the perturbations are fast. However, the same benchmarks
have shown that, for fast fields and not very strong interactions or slow fields irrespective of the interactions, these shortcomings appear to be much less important, and ALDA can give reliable results.

To simulate a disordered 50\% binary alloy, we used a single disorder configuration chosen according to the {\it special quasi-random structure} approach, to effectively describe the random arrangements of sites at short range \cite{Zunger}. This still provides significant insight in localisation trends, while avoiding demanding numerical averaging over 
many configurations \cite{Verdozzi97}. As an indicator of de/localisation, we used a modified inverse participation 
ratio $\zeta$:
\begin{eqnarray}
\zeta=\frac{\sum_i n_{}^2(i)}{\left[\sum_i n_{}(i)\right]^2},
\end{eqnarray}
 exactly accessible within lattice (TD)DFT.

Finally, we examined the KS currents. In TDDFT, the meaning of the KS current-density ${\bf j}_{KS}({\bf r}, t)$ is rather indirect.
It is in fact its divergence  which, via the continuity equation, can be exactly determined (and thus the exact current out a region can be obtained).
The same holds in lattice TDDFT:  rigorous physical content should not be assigned to the KS bond currents 
\begin{eqnarray}
 j^{bond}_{KS}(m;n) = -4 \sum_{k \in occ} \Im \left[\psi_k (m) \overline{\psi_k (n)}\right],
\end{eqnarray}
but rather to their divergence on the lattice ($m,n$ label the nearest neighbour sites defining the bond $mn$). Even so, it may be 
instructive  to consider $ j^{bond}_{KS}$ as an auxiliary, albeit non-rigorous tool to illustrate the dynamics.

\section*{addendum}
\noindent {\it Acknowledgments.} We thank C.-O. Almbladh and J. Hein for valuable discussions.
We gratefully acknowledge LUNARC for computational resources and technical support.
A.K., D.K. and C.V. thank the EOARD (grant FA8655-08-1-3019) and the ETSF (INFRA-2007-211956) for 
financial support.  A.P. thanks M. Capone for the financial support by the European Research Council 
under FP7/ERC Starting Independent Research Grant ÒSUPERBADÓ (Grant Agreement n. 240524). \newline \\
{\it Author contributions.} A.K. developed the computer codes to perform the numerical simulations, 
with contributions from C.V. and D.K. The numerical calculations were 
performed by A.K. The data analysis was performed by A.P. and D.K., with
contributions by C.V. All the authors participated to the discussion of the results and to the
writing of the paper. The project was supervised by C.V.\newline \\
{\it Competing interests statement.} The authors declare that they have no competing financial interests.\newline\\

\section*{The Authors}
\small
\noindent {\it Alexey Kartsev},
Mathematical Physics and European Theoretical Spectroscopy Facility (ETSF), Lund University, 22100 Lund, Sweden\\

\noindent {\it Daniel Karlsson}, Mathematical Physics and European Theoretical Spectroscopy Facility (ETSF), Lund University, 22100 Lund, Sweden\\

\noindent {\it Antonio Privitera}, Democritos National Simulation Center, Consiglio Nazionale delle Ricerche, Istituto Officina dei Materiali (IOM) and Scuola Internazionale Superiore di Studi Avanzati (SISSA), Via Bonomea 265, 34136 Trieste, Italy\\

\noindent {\it Claudio Verdozzi}, Mathematical Physics and European Theoretical Spectroscopy Facility (ETSF), Lund University, 22100 Lund, Sweden ({\bf Corresponding author}: cv@teorfys.lu.se)\\

\end{document}